\documentclass[preprint, 1p, authoryear]{elsarticle}
\usepackage{hyperref}
\hypersetup{pdfauthor=author}
\usepackage{graphicx}
\usepackage{amsmath}
\usepackage{amssymb}
\usepackage{bm}
\usepackage{mathtools}
\usepackage{xcolor}
\usepackage{bbm}

\usepackage{epstopdf}
\usepackage{epsfig}
\usepackage{caption}
\usepackage{subcaption}
\usepackage{afterpage}
\usepackage{amssymb}
\usepackage{stmaryrd}
\allowdisplaybreaks
\makeatletter
\def\ps@pprintTitle{
	\let\@oddhead\@empty
	\let\@evenhead\@empty
	\def\@oddfoot{\centerline{\thepage}}
	\let\@evenfoot\@oddfoot}
\makeatother

\begin{document}
\begin{frontmatter}

\date{\today}
\title{Prior selection for the precision parameter of Dirichlet Process Mixtures}
\author[1]{C.~Vicentini}
\ead{carlo.vicentini@mathstat.net}
\affiliation[1]{organization={Department of Mathematical Sciences, Durham University},
addressline={Upper Mountjoy, Stockton Road},
postcode={DH1 3LE},
city={Durham},
country={UK}}
\cortext[cor1]{Corresponding author}

\author[1]{I.~H.~Jermyn}
\ead{i.h.jermyn@durham.ac.uk}

\begin{abstract}
Consider a Dirichlet process mixture model (DPM) with random precision parameter $\alpha$, inducing $K_n$ clusters over $n$ observations through its latent random partition. Our goal is to specify the prior distribution $p\left(\alpha\mid\boldsymbol{\eta}\right)$, including its fixed parameter vector $\boldsymbol{\eta}$, in a way that is meaningful. 

Existing approaches can be broadly categorised into three groups. Those in the first group depend on the sample size $n$, and often rely on the linkage between $p\left(\alpha\mid\boldsymbol{\eta}\right)$ and $p\left(K_n\right)$ to draw conclusions on how to best choose $\boldsymbol{\eta}$ to reflect one's prior knowledge of $K_{n}$; we call them \textit{sample-size-dependent}. Those in the second and third group consist instead of using quasi-degenerate or improper priors, respectively. 

In this article, we show how all three methods have limitations, especially for large $n$. Then we propose an alternative methodology which does not depend on $K_n$ or on the size of the available sample, but rather on the relationship between the largest stick lengths in the stick-breaking construction of the DPM; and which reflects those prior beliefs in $p\left(\alpha\mid\boldsymbol{\eta}\right)$. We conclude with an example where existing sample-size-dependent approaches fail, while our sample-size-independent approach continues to be  feasible. 	
\end{abstract}
\end{frontmatter}	

\section{Introduction}
Typical usages of the Dirichlet process mixture model (DPM) are density and cluster estimation; the former is motivated by the flexibility of the DPM, and the latter by the latent random partition that the model induces on the observed data. In both cases, the precision parameter $\alpha$ of the DPM is of great significance, since it influences the smoothness of the resulting density, as well as $K_n$, the random number of clusters in the underlying partition of $n$ data points \citep{dorazio_selecting_2009,murugiah_selecting_2012}. 

Methods have been developed to infer $\alpha$ from the data. For example, point estimates can be obtained with empirical Bayes, as outlined in  \cite{liu1996nonparametric}, or alternatively a fully Bayesian approach can be used to obtain the full posterior distribution. This paper focuses on the latter, which often involves Markov Chain Monte Carlo (MCMC); we address the question of how best to choose the parameter vector $\boldsymbol{\eta}$ of the prior distribution $p\left(\alpha\mid\boldsymbol{\eta}\right)$, for some choice of parameterized model, and in particular what quantity of interest one's prior belief should be anchored to when eliciting $\boldsymbol{\eta}$. 

Existing approaches in this domain can be broadly categorised into three groups.  Those in the first group depend on the sample size $n$, and often rely on the linkage between $p\left(\alpha\mid\boldsymbol{\eta}\right)$ and $p\left(K_n\right)$ to draw conclusions on how best to choose $\boldsymbol{\eta}$ to reflect one's prior knowledge of $K_n$; we call them \textit{sample-size-dependent}. Those in the second group consist of quasi-degenerate priors, such as for example a $\textrm{Gamma}\left(a,b\right)$ with $a$ and $b$ both close to zero. Those in the third group consist of improper priors. 

Dependence on the sample size is typically undesirable because the resulting modelling construct, including the data generating process, becomes applicable only for that particular sample size. For example, a common trait of most sample-size-dependent priors is that they try to induce a diffuse prior on $K_n$ \citep{west_hierarchical_1993, dorazio_selecting_2009, murugiah_selecting_2012}. However, a prior deemed diffuse or weakly informative on $K_n$ for a certain sample size $n$ would not necessarily be so for a different value of \textit{n}. In fact, articulating one's prior beliefs through $K_n$ leads to a sequence of priors indexed by the sample size, $\left\{p\left(\alpha\mid\boldsymbol{\eta}_n\right)\right\}_{n=1}^\infty$. This is particularly unappealing because the DPM model is otherwise well suited for inference on streaming data, as the number of clusters that it induces is not fixed, but grows with the data size. Furthermore, by trying to be diffuse in $K_n$, such priors actually influence other important quantities of interest, in a way that is material. In our view, quasi-degenerate and improper priors do not offer viable solutions either, as the former are not compatible with the notion of multiple clusters, hence they defeat the purpose of using a DPM in the first place, while the latter can lead to an improper posterior.

We introduce a new approach to the specification of $p\left(\alpha\right)$, which is independent of the sample size and which is instead based on the appraisal of the implied joint distribution of the stick-breaking weights (either in size-biased order, or ranked); we show an example in which multiple DPMs stem from a common prior $p\left(\alpha;\boldsymbol{\eta}\right)$, and where, as a result, sample-size-dependent approaches are inapplicable while our sample-size-independent method is feasible. Our approach is essentially based on the weights, or lengths, of the sticks that are broken off the initial stick of length $1$, in the stick-breaking representation of the Dirichlet process; these can also be equivalently seen as the asymptotic relative cluster sizes for $n\rightarrow\infty$. 

In the next sections, we proceed as follows. In section \ref{section:dp}, we summarise some basic notions about the Dirichlet process. In section \ref{section:3}, we discuss the existing literature on $\alpha$ priors, and their limitations. In section \ref{section:SSI}, we introduce a novel approach to the selection of $p\left(\alpha\mid\boldsymbol{\eta}\right)$, which does not depend on the sample size. In section \ref{section:comparison}, we cross-examine how sample-size-dependent and sample-size-independent approaches perform in relation to each other. In section \ref{section:5}, we showcase an example where existing sample-size-dependent approaches are inapplicable, while our sample-size-independent approach continues to be  feasible.  Section \ref{section:conclusions} outlines our conclusions.

\section{Representations and properties of the Dirichlet process}\label{section:dp}
Consider a Dirichlet process $G\sim\textrm{DP}\left(\alpha,G_0\right)$ with precision parameter $\alpha$ and base measure $G_0$. As proved in  \cite{ferguson1973}, its posterior distribution given $n$ observations $\left(\theta_1,\ldots,\theta_n\right)$ is in turn a $\textrm{DP}\left(\alpha+n,\frac{\alpha G_0+\sum_{i=1}^{n}\delta_{\theta_i}}{\alpha+n}\right)$ meaning that, unconditional on $G$, we have:
\begin{equation}
    \label{eq:ferguson2}
    p\left(\theta_{n+1}\mid\theta_1,\ldots,\theta_n\right)
    =
    \frac{\alpha}{\alpha+n} G_0\left(\theta_{n+1}\right)
    + 
    \sum_{i=1}^{n} \frac{1}{\alpha+n}\delta_{\theta_i}\left(\theta_{n+1}\right).
\end{equation}

Mixing the DP with the parametric likelihood $p\left(\cdot\mid\theta\right)$ leads to a Dirichlet process mixture (DPM) \citep{lo1984, ferguson_bayesian_1983}, which can be written as
\begin{align}\label{eq:DPMb}
    &y_i\mid \theta_i \sim p\left(y_i\mid\theta_i\right),\\
    &\theta_i\mid G \sim G,\nonumber\\
    &G\sim \textrm{DP}\left(\alpha,G_0\right),\nonumber
\end{align}
where $\boldsymbol{y}=\left(y_1,\ldots,y_n\right)$ is a vector of $n$ observations. 

We discuss two constructions of the Dirichlet process that are relevant to this article: the stick-breaking and the Poisson-Dirichlet process representations.

\subsection{The stick-breaking representation}
\cite{sethuraman} considered 
\begin{align}\label{eq:sethuraman}
	&w_1:=v_1,\nonumber\\
	&w_h:=v_h\prod_{l<h}\left(1-v_l\right),\quad h=2,3,\ldots\nonumber\\
	&v_h\sim \textrm{Beta}\left(1,\alpha\right), \quad h=1,2,\ldots,\\
	&m_h \sim G_0, \quad h=1,2,\ldots,\nonumber
\end{align}
where $G_{0}$ is a probability measure, and proved that the distribution of the random measure
\begin{equation}\label{eq:stick}
	G:=\sum_{h=1}^{\infty}w_h \delta_{m_h}
\end{equation}
is $DP\left(\alpha,G_0\right)$. 

The distribution of $\boldsymbol{w}:=\left(w_1,w_2,\ldots\right)$ alone is known as the Griffiths-Engen-McCloskey distribution, and it is denoted by $\textrm{GEM}\left(\alpha\right)$ \citep[section 4.8]{arratia_logarithmic_2003}; its elements $w_1, w_2, \ldots$ are in size-biased order \citep{pitman_random_1996}. The probability distribution of the first $H-1$ elements of $\boldsymbol{w}$ can be derived from the $H$-dimensional generalised Dirichlet distribution \citep{connor1969concepts}.

\subsection{The Poisson-Dirichlet Process representation}\label{section:pdd}
\cite{kingman_random_1975} introduced the Poisson-Dirichlet distribution (PDD), which he constructed through the gamma process $\xi\left(t\right)$, a stochastic process with $\xi\left(0\right)=0$ and with increments which are independent on disjoint intervals, and gamma distributed. The PDD with parameter $\alpha$ is known to be equivalent to the distribution of the decreasing order statistics of $\boldsymbol{w}$ \citep[section 4.11]{arratia_logarithmic_2003}\citep{pitman_random_1996}; we denote them by $\left(w_1^\shortdownarrow,w_2^\shortdownarrow,\ldots\right)$. 
Similarly, the decreasing weight-ranked equivalent of equation \ref{eq:stick} is known as the Poisson-Dirichlet Process (PDP).

As laid out in \cite{watterson1976stationary}, the conditional joint probability distribution of the first $r$ ranked weights is:
\begin{equation}\label{eq:watterson}
	p\left(w_1^\shortdownarrow,\ldots, w_r^\shortdownarrow\right)=\alpha^r \Gamma\left(\alpha\right) e^{\gamma\alpha}\frac{w_r^{\shortdownarrow \alpha-1}}{w_1^\shortdownarrow\cdots w_r^\shortdownarrow}\ g\left(\frac{1-w_1^\shortdownarrow-\ldots-w_r^\shortdownarrow}{w_r^\shortdownarrow}\right),
\end{equation}
where $g$ is a function which can be recursively written as:
\begin{equation*}
	g\left(z\right)=z^{\alpha-1}\left[g\left(n\right)n^{1-\alpha}-\alpha\int_n^z g\left(y-1\right)y^{-\alpha}dy\right], \quad n\leq z<n+1,\ n:=\lfloor z\rfloor,
\end{equation*}
and which is known to be particularly difficult to compute.

\subsection{Properties of the Dirichlet process}\label{section:2}
Repeated values in $\boldsymbol{\theta}$ induce implicit clustering on $\boldsymbol{\theta}$ and, in turn, a random partition model. This can be observed in equation \ref{eq:ferguson2}, where if $G_0$ is continuous, we obtain
\begin{equation*}
	p\left(\theta_i\notin\left\{\theta_1,\ldots,\theta_{i-1}\right\}\mid\alpha\right)=\frac{\alpha}{\alpha+i-1}, \ i=2,\ldots,n,
\end{equation*}
hence the random number of clusters $K_n\mid\alpha$ observed in a sample of $n$ observations is distributed as the sum of $n$ Bernoulli variables with parameters $p_i=\alpha/\left(\alpha+i-1\right)$. Further, 
\begin{align}
	\mathbb{E}\left[K_n\mid\alpha\right]&=\sum_{i=1}^n\frac{\alpha}{\alpha+i-1}\sim \alpha\log\frac{n+\alpha}{\alpha}\sim \alpha \log n,\label{eq:K:mean}\\
	\mathbb{V}\textrm{ar}\left[K_n\mid\alpha\right]&=\sum_{i=1}^n\frac{\alpha\left(i-1\right)}{\left(\alpha+i-1\right)^2}\sim \alpha \log n,\label{eq:K:variance}
\end{align}
as $n\rightarrow\infty$ \citep{antoniak, arratia_logarithmic_2003}. 
The probability distribution of $K_n\mid\alpha$ is:
\begin{equation}\label{eq:K}
p\left(K_n=k\mid\alpha\right)=s_{n,k} \ \alpha^k \ \frac{\Gamma\left(\alpha\right)}{\Gamma\left(\alpha+n\right)},
\end{equation}
where $s_{n,k}$ are unsigned Stirling numbers of the first kind. The parameter $\alpha$ therefore critically controls $p\left(K_n\mid\alpha\right)$: we show in sections \ref{section:degenerate} and \ref{section:improper2} that, for $\alpha\rightarrow 0$, the random variable $K_n\mid\alpha$ converges to the Dirac measure $\delta_1$, while for $\alpha \rightarrow \infty$, the same converges to $\delta_n$; similarly, if $\alpha$ is random, $p\left(\alpha\mid\boldsymbol{\eta}\right)$ determines the prior  $p\left(K_n\right)$ through mixing, as
\begin{equation}\label{eq:Kn}
p\left(K_n=k\right)=\int_0^{\infty} s_{n,k} \ \alpha^k \ \frac{\Gamma\left(\alpha\right)}{\Gamma\left(\alpha+n\right)} \textrm{d}p\left(\alpha\mid\boldsymbol{\eta}\right).
\end{equation}
The following also holds \citep{watterson_sampling_1974, arratia_number_2000}:
\begin{equation}
d_{TV}\left(p\left(K_n\mid\alpha\right),\textrm{Po}\left(\mathbb{E}\left[K_n\mid\alpha\right]\right)\right)=O\left(\frac{1}{\log n}\right),\label{equation:poiConvergence}
\end{equation}
where $d_{TV}$ indicates total variation distance and $\textrm{Po}\left(\lambda\right)$ indicates the Poisson distribution with parameter $\lambda$, and which implies convergence in distribution since $1/\log n\rightarrow 0$ for $n\rightarrow\infty$. While the relationship above is unlikely to carry practical use in computations, as its rate of convergence is sublinear, it does provide conceptual insight into the limiting behaviour of $p\left(K_n\mid\alpha\right)$, as we show in section \ref{sec:ssd}.

\section{Existing priors for \texorpdfstring{$\alpha$}{alpha}}\label{section:3}

In this section we discuss three approaches to prior specification. Methods in the first group (see section \ref{sec:ssd}) are \textit{sample-size-dependent}, and we refer to them as `SSD'. Those in the other two groups are quasi-degenerate and improper priors (see sections \ref{section:degenerate} and \ref{section:improper2}).

Although in principle any distributional choice of $\alpha$ is allowed, all aforementioned approaches have historically been discussed by their authors in the context of the gamma distribution -- i.e. $\alpha\sim\textrm{Ga}\left(a,b\right)$. The popularity of the gamma distribution as a prior for $\alpha$ is due to reasons of computational attractiveness of the posterior, and is to be traced back to \cite{escobar1995bayesian}.

\subsection{Sample-size-dependent approaches (SSD)}\label{sec:ssd}
Methods in this group are the $K_n$-diffuse prior of \cite{west_hierarchical_1993}, the DORO prior of \cite{dorazio_selecting_2009}, the SCAL prior of \cite{murugiah_selecting_2012}, and Jeffreys' prior \citep{rodriguez2013jeffreys}. The first three leverage one's prior assumptions on $K_n$ as a target to determine $\boldsymbol{\eta}$ in $p\left(\alpha\mid\boldsymbol{\eta}\right)$, while Jeffreys' prior purely depends on $n$ and does not involve one's assumptions on $K_n$. All four depend on the size of the sample, $n$. 

\subsubsection{\texorpdfstring{$K_n$-diffuse prior}{K-diffuse priors}}\label{sec:diffuse}

\cite{west_hierarchical_1993} use a gamma hyperprior, $\alpha\sim\textrm{Ga}\left(a,b\right)$, ``supporting a diffuse range of reasonably large values consistent with possibly large values'' of $K_n$. In their article, they use $n=74$; their prior supports ``a wide range of $k$ values between about $k=8$ and $k = 35$''\footnote{\cite{west_hierarchical_1993} use $a=5$ and $b=0.5$, and $k$ in their notation is equivalent to $K_n$ in ours.}. We call their approach $K_n$-diffuse, to highlight that it is not necessarily diffuse in $\alpha$, but rather it is diffuse in $K_n$.

\begin{figure}[htb]
	\centering
	\includegraphics[width=0.98\textwidth]{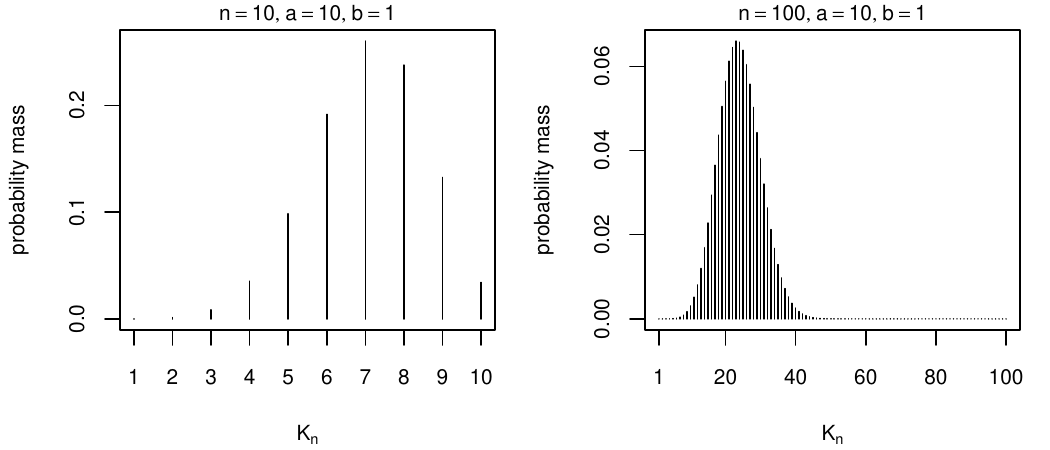}
	\caption{$K_n$-diffuse prior. While the prior probability distribution induced on $K_n$ by $\alpha\sim\textrm{Ga}\left(10,1\right)$ appears reasonably diffuse for $n=10$, it is less so for $n=100$, as the shape and skewness of $p\left(K_n\right)$ change with $n$.}
	\label{figure:diffusePrior}
\end{figure}

We observe that this approach, which is appealing in its simplicity, has a dependency on the size of the sample it is applied to. For example, Figure \ref{figure:diffusePrior} shows the impact on $p\left(K_n\right)$  of a $\textrm{Ga}\left(10,1\right)$ prior, as $n$ moves from $n=10$ to $n=100$:
\begin{itemize}
	\item in the left panel ($n=10$), $K_n$ is centred on values that are large relative to $n$, with a wide spread relative to the support of $K_n$, hence it is well-diffused;
	\item  in the right panel ($n=100$), $K_n$ is centred on smaller values (relative to $n$), with a smaller spread over the support of $K_n$ -- meaning that it is not as well-diffused as when $n=10$.
\end{itemize}
As a result, we conclude that if a $K_n$-diffuse prior is defined as one whose mass is well-spread around central values of $K_n$, the consequence is that, under a $\textrm{Ga}\left(a,b\right)$ prior, $\left(a, b\right)$ needs to be updated as $n$ grows, to ensure that central values of $K_n$ continue to be well covered, and that the relative spread around central values is preserved. 

\subsubsection{DORO priors}\label{section:dorazio}
\cite{dorazio_selecting_2009} proposes $\alpha\sim \textrm{Ga}\left(a,b\right)$, with $\left(a,b\right)$ set to minimise the Kullback-Leibler distance between the prior probability distribution $p\left(K_n\right)$ induced by $p\left(\alpha\mid\boldsymbol{\eta}\right)$ on $K_n$, and a target discrete distribution; in the absence of prior information about $K_n$, \cite{dorazio_selecting_2009} uses the discrete uniform as a target. The DORO approach results in a pre-determined list of optimal values of $\left(a,b\right)$ for various values of $n$ (Table \ref{table:doro}).

\begin{figure}[htb]
	\centering
	\includegraphics[width=0.98\textwidth]{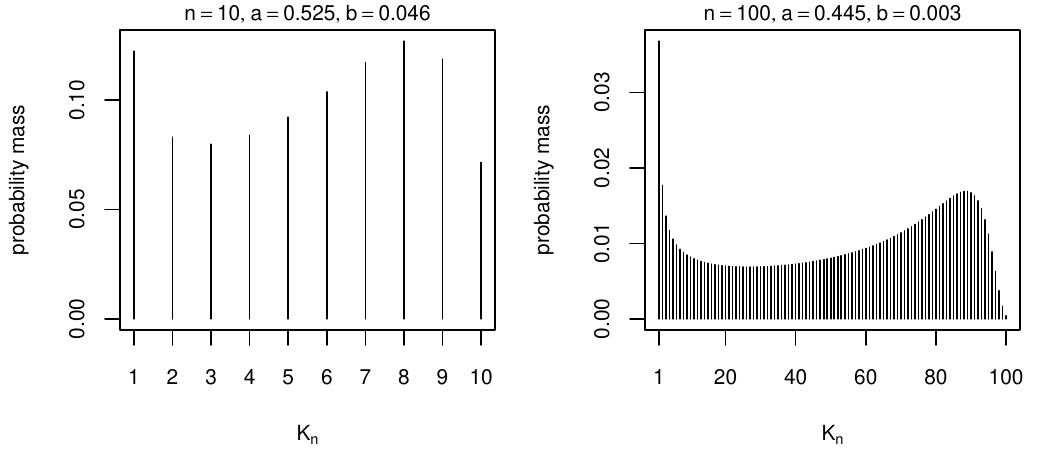}
	\caption{DORO prior. Prior probability distribution induced on $K_n$ by $\alpha\sim\textrm{Ga}\left(a,b\right)$. $K_n$ does not appear to be close to the target discrete uniform distribution, and the approximation does not appear to improve as $n$ increases.}
	\label{figure:dorazio}
\end{figure}

However, Figure \ref{figure:dorazio} shows that the approximation of the discrete uniform resulting from the DORO prior is visibly coarse, and that it does not appear to improve as $n$ grows. In fact, asymptotic results show that, when $\alpha\sim\textrm{Ga}\left(a,b\right)$, $K_n$ converges sub-linearly to a negative binomial random variable, meaning that a discrete uniform is unachievable for $n \rightarrow \infty$: from equation \ref{equation:poiConvergence}, 
\begin{equation*}
	p\left(K_n\mid\alpha\right)\xrightarrow[]{d}\textrm{Poi}\left(\alpha\log n\right), \quad n\to\infty,
\end{equation*}
and $\alpha\log n \sim \textrm{Ga}\left(a, b/\log n\right)$ hence $K_n\sim\textrm{NB}\left(a,b/\left(b+\log n\right)\right)$ in the limit.

Furthermore, as we will see in section \ref{section:jeffreys}, the shape of the prior induced by DORO on $K_n$ is also quite different from the one that is attained under Jeffreys' prior, which is one more reason why the choice by DORO of targeting a discrete uniform prior distribution on $K_n$ can be debated. 

\begin{table}[ht]
	\centering
	\begin{tabular}{rrrr}
		$n$ & $a$ & $b$ & $D_{KL}$\\ \hline
		5	& 0.541	& 0.096 & 0.00458\\
		10	& 0.525 & 0.046 & 0.01904\\
		15	& 0.512 & 0.029 & 0.03048\\
		20	& 0.501 & 0.021 & 0.03942\\
		25	& 0.490 & 0.015 & 0.04660\\
		30	& 0.486 & 0.013 & 0.05272\\
		35	& 0.480 & 0.010 & 0.05806\\
		40	& 0.475 & 0.009 & 0.06265\\
		45	& 0.470 & 0.008 & 0.06684\\
		50	& 0.467 & 0.007 & 0.07050\\
		100 & 0.445 & 0.003 & 0.09529\\
	\end{tabular}
	\caption{Optimal values of $a,b$ under the DORO approach, when $\alpha\sim\textrm{Ga}\left(a,b\right)$ and when the target distribution $p\left(K_n\right)$ is the discrete uniform. We have enriched the original table from \cite{dorazio_selecting_2009} with an additional entry for $n=100$.}
	\label{table:doro}	
\end{table}

\subsubsection{SCAL priors}\label{section:murugiah}
\cite{murugiah_selecting_2012} propose a scaling approach, where the values of $\left(a,b\right)$ are initially computed for a given $n$ according to how well they perform at recovering some known cluster structure; $\left(a,b\right)$ are subsequently rescaled to other choices of $n$, without the need to re-compute them again through the fully-fledged determination process. 

Upon the initial determination of $\left(a,b\right)$, the goal of SCAL is to scale $\left(a,b\right)$ in such a way that the prior mean of $K_n$ is only affected to a small extent by the changes, while the variance is influenced to a larger extent. In particular, SCAL is based on fixing $p\left(K_n=1\right)$ and $p\left(K_n \in \{c, c+1, \ldots, n-1, n\}\right)$ to some determined values, for a suitably selected $c$, and deriving $\left(a,b\right)$ implicitly as $n$ varies, with equation \ref{eq:Kn}. \cite{murugiah_selecting_2012} suggest $c= \lceil c_0\log n \rceil$, with $c_0=2$. 

\cite{murugiah_selecting_2012} use simulated data sets of $n=6$ observations to elicit $\left(a,b\right)=\left(1,1\right)$; they then test their scaling approach up to $n=25$, by re-calculating $\left(a,b\right)$ while keeping $p\left(K_n=1\right)=0.34$ and $p\left(K_n \in \left\{c,\ldots,n\right\}\right)=0.15$. They fit a curve through the results that they obtain, to ultimately propose the following equation, as an easier  approximation\footnote{This is justified in \cite{murugiah_selecting_2012} by observing that $\mathbb{E}\left[\alpha\right]=1$, which is fixed, and that $\mathbb{V}\left[\alpha\right]=e^{0.0165 n}$, which increases with $n$, as their approach originally intended.} to SCAL:
\begin{equation}\label{eq:scaling}
	a=b=e^{-0.033 n}.
\end{equation}

Although \cite{murugiah_selecting_2012} only obtained the optimal $\left(a,b\right)$ values for their test case for $n \in \left\{6,10,15,20,25\right\}$, we extend their results to $n \in \left\{50,75,100\right\}$ (see table \ref{table:scal}). As expected, we observe dependence on $n$, albeit to a lesser extent than DORO. We also note that the proposed approximation significantly diverges from the exact values for $n>25$; in fact, instead of the exact values $\left(a,b\right)=\left(0.403,0.370\right)$, for $n=100$ it yields $a=b=0.037$, a quasi-degenerate gamma prior.

\begin{table}[ht]
	\centering
	\begin{tabular}{ccc}
		$n$ & exact $\left(a,b\right)$ & approx. $\left(a,b\right)$\\ \hline
		25	& $\left(0.490,0.438\right)$ & $\left(0.438,0.438\right)$\\
		50	& $\left(0.466,0.467\right)$ & $\left(0.192,0.192\right)$\\
		75	& $\left(0.432,0.420\right)$ & $\left(0.084,0.084\right)$\\
		100	& $\left(0.403,0.370\right)$ & $\left(0.037,0.037\right)$\\
	\end{tabular}
	\caption{Optimal and approximate values of $\left(a,b\right)$ under the SCAL approach, as we determined them to be according to the approach outlined in \cite{murugiah_selecting_2012}. Approximate values originate from equation \ref{eq:scaling}. Targets are $p\left(K_n=1\right)=0.34$ and $p\left(K_n \in \left\{c,\ldots,n\right\}\right)=0.15$, $c=c_0\log n$, $c_0=2$.}
	\label{table:scal}	
\end{table}

\subsubsection{Jeffreys' prior}\label{section:jeffreys}
Jeffreys'  prior was introduced by \cite{jeffreys1946invariant} as the volume measure of the Riemannian metric induced on the parameter space from a Riemannian metric on the space of probability distributions containing the parameterized model, thereby ensuring that the prior is determined by the associated probability laws and not by an arbitrary labelling of those laws by parameter values. This would be true for any Riemannian metric on the space of distributions; what makes Jeffreys' prior special is that it is induced by the unique \citep{Cencov1982-ob,Campbell1986-qi,Bauer2016-jm} Riemannian metric that is invariant to (appropriately behaved) mappings of the space upon which the distribution is defined. Subsequently, it was found to have other desirable properties too, and it is now widely used in statistics and best known as the most prominent example of an \textit{objective} prior. It was first derived for the multivariate Ewens distribution by \cite{rodriguez2013jeffreys}, who then tested it on the related Dirichlet process mixture model.

For univariate cases like the one being discussed, it is obtained as:
\begin{equation*}
p\left(\alpha\right)\propto\sqrt{I_\alpha\left(\alpha\right)}=\sqrt{\mathbb{E}_k\left[\left(\frac{\partial}{\partial\alpha}\log p\left(k\mid\alpha\right)\right)^2\right]}=\sqrt{\frac{1}{\alpha}\sum_{i=1}^n\frac{i-1}{\left(\alpha+i-1\right)^2}},
\end{equation*}
where $I_\alpha\left(\alpha\right)$ is the Fisher information (see \cite{rodriguez2013jeffreys}). We found the following equivalent formulation to be computationally faster:
\begin{equation*}
p\left(\alpha\right)\propto \sqrt{\frac{1}{\alpha}\left[\psi_0\left(\alpha+n\right)-\psi_0\left(\alpha+1\right)+\alpha\left(\psi_1\left(\alpha+n\right)-\psi_1\left(\alpha+1\right)\right)\right]},
\end{equation*}
which is obtained because, by definition, $\psi_0\left(z+1\right)=\psi_0\left(z\right)+\frac{1}{z}$, and also $\psi_1\left(\alpha+1\right)=\psi_1\left(\alpha\right)-\frac{1}{\alpha^2}$.

Its density is $0$ on the entire half-line for $n=1$, as is its integral, hence this prior carries no meaning for $n=1$. This is natural, since when $n = 1$, $k = 1$ with certainty, and there is no dependence on $\alpha$.  

For $n=2$, Jeffreys' prior can be expressed analytically, including its normalising constant, leading to the density and cumulative probability distribution functions:
\begin{align}
&p\left(\alpha\right)=\frac{1}{\pi\left(\alpha+1\right)\sqrt{\alpha}},\label{eq:Jdensity}\\
&p\left(\alpha\leq x\right)=\frac{2}{\pi}\textrm{atan}{\sqrt{\alpha}}\label{eq:Jcumulate}.
\end{align}

\begin{figure}[htb]
\centering
\includegraphics[width=0.98\textwidth]{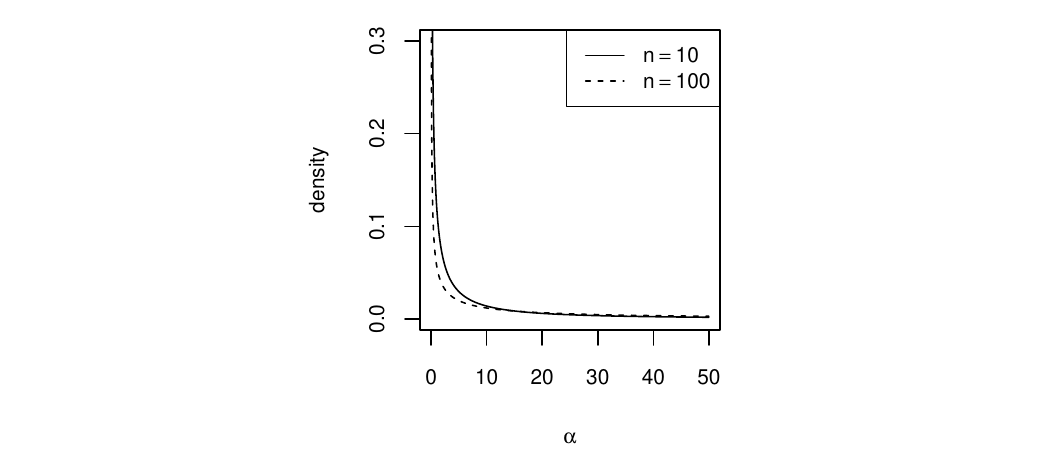}
\caption{Density of Jeffreys' prior for $n=10$ and $n=100$.}
\label{figure:jeffreys-plain}
\end{figure}

This prior is proper and has no finite moments \citep{rodriguez2013jeffreys}. Figure \ref{figure:jeffreys-plain} exemplifies the change in shape of $p\left(\alpha\right)$ as $n$ increases.

\begin{figure}[htb]
\centering
\includegraphics[width=0.98\textwidth]{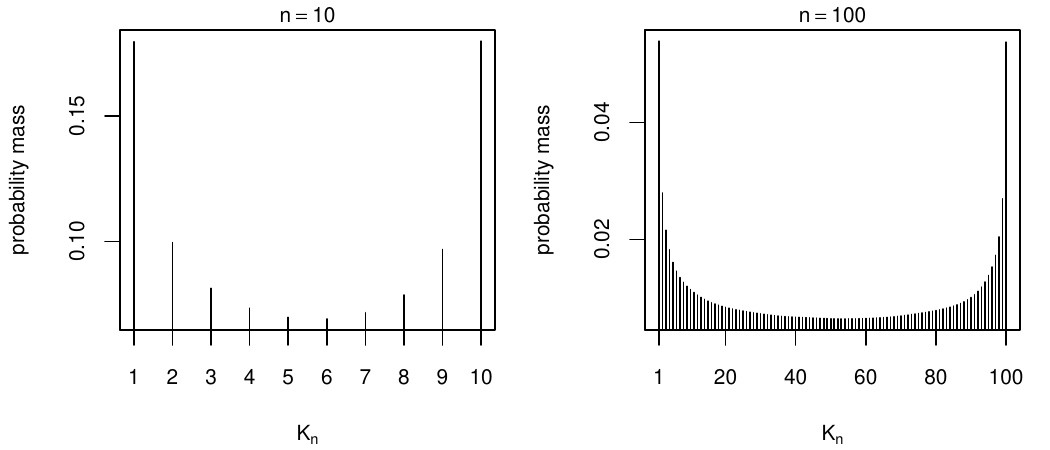}
\caption{Prior distribution $p\left(K_n\right)$ induced by assigning Jeffreys' prior to $p\left(\alpha\right)$, for $n=10$ and $n=100$.}
\label{figure:jeffreys-k}
\end{figure}

Two examples of the shape of the prior distribution induced on $K_n$ by assigning Jeffreys' prior to $\alpha$ are plotted in figure \ref{figure:jeffreys-k}. As noted in \cite{rodriguez2013jeffreys},  this distribution seems to be approximately symmetric around $n/2$. We include in \ref{appendix:proprietyA} and \ref{appendix:proprietyK}  results showing that the posterior distribution induced by Jeffreys' prior on $\alpha$ is proper, as is the prior distribution it induces on $K_n$.

In their paper, after testing it on a species sampling model, \cite{rodriguez2013jeffreys} sample $50$ data points from a negative binomial $\textrm{NB}\left(20,220\right)$, and fit a DPM with a Gamma base measure (with mean and variance to mirror that of the negative binomial) under two priors for $\alpha$: the quasi-degenerate $\textrm{Ga}\left(0.001,0.001\right)$ prior, and Jeffreys' prior. They find that posterior inference on $K_n$ with the DPM leads to very similar results under these two options, with Jeffreys' prior favoring "a somewhat larger number of clusters than the Gamma prior" \citep{rodriguez2013jeffreys}. They also stress that Jeffreys' prior "explicitly depends on the sample size".

\subsection{Quasi-degenerate priors}\label{section:degenerate}
Similarly to the SCAL approximation from equation \ref{eq:scaling}, other authors have independently suggested, on isolated occasions rather than as part of an attempt to design a prior elicitation framework, to use a quasi-degenerate  $\textrm{Ga}\left(a,b\right)$, with $a$ and $b$ close to zero, on the basis that it approximates the improper prior $p\left(\alpha\right)\propto 1/\alpha$ that is uniform in $\log\left(\alpha\right)$ \citep{escobar1995bayesian, escobar1998computing,navarro_modeling_2006,bugs}. However, gamma priors with very small $\left(a,b\right)$ are  undesirable, because
\begin{equation*}
	\lim_{\left(a,b\right)\to \left(0,0\right)}p\left(K_n=1\right)=1,
\end{equation*}
hence they lead to the prior expectation of a parametric model with only one mixture component, which negates the reason for using a nonparametric prior in the first place and which is likely to overwhelm the data due to its strength; \cite{dorazio_selecting_2009} and \cite{murugiah_selecting_2012} also draw similar conclusions. 

This is proven as follows. We are interested in
\begin{equation}\label{eq:pk}
	\lim_{\left(a,b\right)\to \left(0,0\right)} p\left(K_n=k\right)=s_{n,k}\lim_{\left(a,b\right)\to \left(0,0\right)}\int_{0}^{\infty}\alpha^k\frac{\Gamma\left(\alpha\right)}{\Gamma\left(\alpha+n\right)}\textrm{d}p\left(\alpha\mid a,b\right),
\end{equation}
where $p\left(\alpha\mid a,b\right)$ is the gamma probability measure with parameters $\left(a,b\right)$. In what follows, we re-write equation \ref{eq:pk} as the limit of a sequence, and we use results on the weak convergence of measures to prove that the sequence converges to $1$ for $k=1$, and to $0$ for every other admissible value of $k$.

For each path in $\mathbb{R}^+ \times \mathbb{R}^+$ where $\left(a,b\right)\to \left(0,0\right)$, we define $\left\{X_l\right\}$, a sequence of gamma distributed random variables $X_1, X_2, \ldots$, each with parameters $\left(a_1,b_1\right), \left(a_2,b_2\right), \ldots$ identified along that path. The distribution function of $X_l$ is
\begin{equation*}
	F_{X_l}\left(\alpha;a_l,b_l\right)=\frac{\gamma\left(a_l,b_l\alpha\right)}{\Gamma\left(a_l\right)},
\end{equation*}
where $\gamma\left(\right)$ is the lower incomplete gamma function. It is well known that
\begin{equation*}
	\frac{\gamma\left(s,t\right)}{t^s}\to\frac{1}{s},
\end{equation*}
as $t\to 0$. Hence
\begin{align*}
	\lim_{\left(a,b\right)\to\left(0,0\right)}\gamma\left(a,b\alpha\right)\frac{1}{\Gamma\left(a\right)}&=\lim_{\left(a,b\right)\to\left(0,0\right)}\frac{\left(b\alpha\right)^{a}}{a}\frac{1}{\Gamma\left(a\right)}=\lim_{l\to\infty}\frac{\left(b_l\alpha\right)^{a_l}}{a_l}\frac{1}{\Gamma\left(a_l\right)}\\
	&=\lim_{l\to\infty}{\left(b_l\alpha\right)^{a_l}}\frac{1}{a_l\Gamma\left(a_l\right)}=\lim_{l\to \infty}{b_l}^{a_l} \cdot \lim_{l \to \infty}\alpha^{a_l}=1, \quad \forall \alpha>0,
\end{align*}
where we use the fact that
\begin{equation*}
	\lim_{\left(a,b\right)\to\left(0,0\right)}b^a=\lim_{r\to 0}\left(r\sin\theta\right)^{r\cos\theta}=\lim_{r\to 0}\left(r^r\right)^{\cos\theta}\left(\sin^r\theta\right)^{\cos\theta}=1,
\end{equation*}
which holds as the paths along $a=0$ and $b=0$ do not belong to the function domain $\mathbb{R}^+ \times \mathbb{R}^+$ of $F_{X_l}$. 

Therefore $X_l \xrightarrow[]{d} 0$, since the distribution function of the r.v. $X=0$ is equal to $1$ over the continuity set $\left(0,\infty\right)$ of $X$, for every path and every sequence $\left\{X_l\right\}$. A consequence is that the sequence of gamma probability measures $\left\{p_l\right\}$ induced by $\left\{X_l\right\}$ converges weakly to the Dirac measure $\delta_0$. Hence equation \ref{eq:pk} becomes
\begin{equation*}
	\lim_{\left(a,b\right)\to \left(0,0\right)} p\left(K_n=k\right)=s_{n,k}\int_{0}^{\infty}\alpha^k\frac{\Gamma\left(\alpha\right)}{\Gamma\left(\alpha+n\right)}\textrm{d}\delta_0\left(\alpha\right)=
	\begin{cases}
		1,\quad k=1,\\
		0,\quad k=2,\ldots,n,
	\end{cases}
\end{equation*}
as $\alpha^k\Gamma\left(\alpha\right)/\Gamma\left(\alpha+n\right)$ is bounded and continuous.

\subsection{Improper priors}\label{section:improper2}
In the preceding subsection, we mentioned how some studies motivate the use of quasi-degenerate gamma priors on the basis that they approximate the improper prior $p\left(\alpha\right)\propto 1/\alpha$, therefore implying that improper priors are desirable.

However, using $p\left(\alpha\right)\propto 1/\alpha$ leads to an improper posterior $p\left(\alpha\mid\boldsymbol{y}\right)$ as well as to an improper implied prior $p\left(K_n\right)$, and so does $p\left(\alpha\right)\propto 1$. In fact, from equation \ref{eq:K} 
we obtain
\begin{equation*}
p\left(\alpha\mid K_n=k\right)\propto\alpha^k \frac{\Gamma\left(\alpha\right)}{\Gamma\left(\alpha+n\right)}\cdot\frac{1}{\alpha}\sim \frac{1}{\alpha^{n-k+1}} \quad \textrm{as}\  \alpha\to\infty,
\end{equation*}
which is not integrable on any $\left(t,\infty\right)$ interval in the domain for $k=n$, meaning that the $\alpha$ posterior induced by the prior $1/\alpha$ cannot be normalised and is improper. 

The $p\left(\alpha\right)\propto 1/\alpha$ prior also induces an improper prior on $K_n$, as 
\begin{equation*}
p\left(K_n=k\right)=s_{n,k}\int_0^{\infty} \alpha^k \frac{\Gamma\left(\alpha\right)}{\Gamma\left(\alpha+n\right)}\cdot\frac{1}{\alpha}\:\textrm{d}\alpha,
\end{equation*}
where the integrand diverges as above. 

Similar considerations apply when the $\alpha$ prior is $p\left(\alpha\right)\propto 1$, the conclusion being that the $\alpha$ posterior and the induced $K_n$ prior are improper because they are divergent for $k=n$ and $k=n-1$.

\section{Sample-size-independent priors for \texorpdfstring{$\alpha$}{alpha} (SSI)}\label{section:SSI}
We introduce a new prior selection approach which is independent of sample size, and which is motivated by the fact that, irrespective of how many $K_n$ clusters are observed in a sample of size $n$, there is always an underlying infinite-dimensional collection of point masses induced by the $\textrm{DP}\left(\alpha,G_0\right)$, which is independent of $n$ and which can be made to reflect one's prior beliefs. These point masses are the weights, or lengths, of the sticks that are broken off the initial stick of length $1$, in the stick-breaking representation of the Dirichlet process. They also equivalently correspond to the asymptotic relative cluster sizes for $n\rightarrow\infty$. We consider two options: their size-biased random permutation, and their ranked permutation. We henceforth refer to this approach as SSI (sample-size-independent).

\subsection{Size-biased weights}\label{sec:size}
In the stick-breaking representation of equation \ref{eq:stick}, a DP is a collection of infinitely many point masses. In particular, $\left\{w_h\right\}_{h=1}^\infty$ is a sequence of stochastically decreasing weights; it is said to be a size-biased random permutation of the weights because the probability of each weight $w_i$ being in first position in the size-biased permutation is precisely $w_i$ \citep{pitman_random_1996}. A priori, the probability distribution of the weights as they naturally arise in the stick-breaking construction of equation \ref{eq:sethuraman} and \ref{eq:stick}  is the same as the probability distribution of the size-biased weights, hence for simplicity we do not distinguish between them in notation.

\begin{figure}[htb]
\centering
\includegraphics[width=0.98\textwidth]{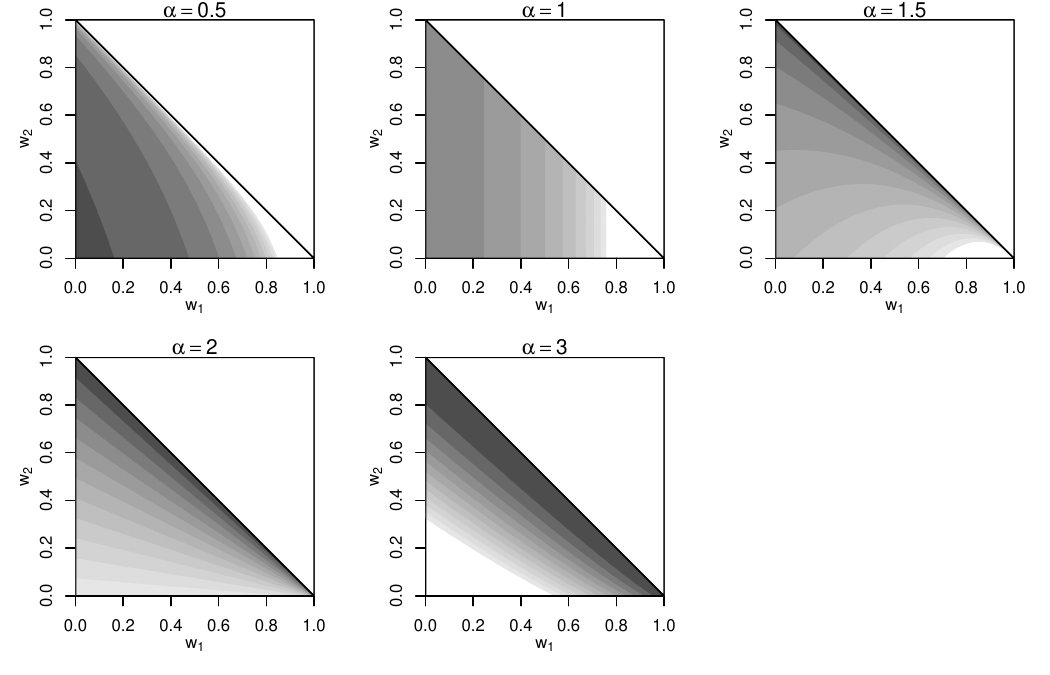}
\caption{Sample-size-independent approach, size-biased. Conditional joint probability distribution $p\left(w_1,w_2\mid\alpha\right)$, for different values of $\alpha$. Darker colours indicate smaller values.}
\label{figure:SB_conditional}
\end{figure}

The conditional joint probability distribution of the first $H$ size-biased weights is a particular case of the generalised Dirichlet distribution \citep{connor1969concepts}:
\begin{equation}\label{eq:connorMosimann}
p\left(w_1,\ldots,w_H\mid\alpha\right)=\alpha^H \ \frac{\left(1-w_1-\ldots-w_H\right)^{\alpha-1}}{\left(1-w_1\right)\ldots\left(1-w_1-\ldots-w_{H-1}\right)}.
\end{equation}
We plot its first two elements in Figure \ref{figure:SB_conditional}, for selected values of $\alpha$. Since $p\left(w_1,w_2,\ldots\right)$ is infinite-dimensional, we restrict our analysis to a finite number of dimensions; for practical purposes and for simplicity, we focus on the bivariate distribution of $w_1$ and $w_2$; our approach can in principle be extended to more dimensions, if necessary. Analytically, we have that:
\begin{equation*}
p\left(w_1,w_2\mid\alpha\right)=\frac{\alpha^2}{1-w_1}\left(1-w_1-w_2\right)^{\alpha-1}.
\end{equation*}

The density $p\left(w_1,w_2\mid\alpha\right)$ attains its maximum on $w_2=1-w_1$ for $\alpha<1$, at $\left(1,0\right)$ for $1\leq\alpha<2$, at $w_2=0$ for $\alpha=2$ and at $\left(0,0\right)$ for $\alpha>2$ (Figure \ref{figure:SB_conditional}). Its partial derivatives are:
\begin{align*}
\frac{\partial p\left(w_1,w_2\mid\alpha\right)}{\partial w_1}&=\alpha^2 \frac{\left(1-w_1-w_2\right)^{\alpha-2}}{\left(1-w_1\right)^2}\left(\left(\alpha-2\right)w_1-w_2-\alpha+2\right),\\
\frac{\partial p\left(w_1,w_2\mid\alpha\right)}{\partial w_2}&=\alpha^2\left(1-\alpha\right)\frac{\left(1-w_1-w_2\right)^{\alpha-2}}{1-w_1},
\end{align*}
and analysis of their sign leads to the considerations in Table \ref{table:derivatives}.

\begin{table}[tb]
\centering
\begin{tabular}{lll}
	$\alpha$ & $p\left(w_1\mid w_2,\alpha\right)$ & $p\left(w_2\mid w_1,\alpha\right)$ \\ \hline
	$0<\alpha<1$ & increasing & increasing\\
	$\alpha=1$ & increasing & constant\\ 
	$1<\alpha<2$ & concave; max. attained at $w_1=1-\frac{w_2}{2-\alpha}$ & decreasing \\ 
	$\alpha=2, w_2\neq 0$ & decreasing & decreasing \\ 
	$\alpha=2, w_2=0$ & constant & decreasing \\ 
	$\alpha>2$ & decreasing & decreasing \\ 
\end{tabular}
\caption{Sample-size-independent approach, size-biased. Behaviour of $w_1\mid w_2,\alpha$ and $w_2\mid w_1,\alpha$, for different values of $\alpha$.}
\label{table:derivatives}	
\end{table}

Our prior belief in the cases displayed in Table \ref{table:derivatives} can be used to inform our choice of the parameter vector $\boldsymbol{\eta}$ of the prior distribution $p\left(\alpha\mid\boldsymbol{\eta}\right)$, since assigning a prior to $\alpha$ means mixing over those cases. 

Denote by $\boldsymbol{\eta}=\left(\eta_1,\ldots,\eta_d\right)$ the $d$ parameters of the continuous probability distribution $p\left(\alpha\mid\boldsymbol{\eta}\right)$, and denote by $p_{0,t_1},\ldots,p_{t_d,\infty}$ our prior belief associated with a partition of $\left(0,\infty\right)$ into $d+1$ nonempty subsets $\left\{\left(0,t_1\right],\ldots,\left(t_{d},\infty\right)\right\}$. For example, when $d=2,\ t_1=1,\ t_2=2$, we partition $\left(0,\infty\right)$ into $\left(0,1\right]$,$\left(1,2\right]$ and $\left(2,\infty\right)$.

We then choose $\boldsymbol{\eta}$ so that $p\left(\alpha\mid\boldsymbol{\eta}\right)$ reflects our prior belief by solving:
\begin{equation*}
\begin{cases}
	p\left(0<\alpha\leq t_1\mid\boldsymbol{\eta}\right)=p_{0,t_1},\\
	p\left(t_1<\alpha\leq t_2\mid\boldsymbol{\eta}\right)=p_{t_1,t_2},\\
	\ldots\\
	p\left(\alpha>t_d\mid\boldsymbol{\eta}\right)=p_{t_d,\infty}.
\end{cases}
\end{equation*}

The resulting system of equations can be either solved analytically, if the cumulative probability distribution admits an explicit representation of its inverse, or numerically. 
For example, this is analytically feasible when $\alpha\sim\textrm{Exp}\left(\eta\right)$, and we obtain:
\begin{equation}\label{eq:explicit}
\eta=\log\left(\left(1-p_{0,t_1}\right)^{-\frac{1}{t_1}}\right),
\end{equation}
and clearly, when $d=1$, $p_{t_1,\infty}=1-p_{0,t_1}$. 

When instead $\alpha\sim\textrm{Ga}\left(\eta_1,\eta_2\right)$, we obtain:
\begin{equation*}
\begin{dcases}
	\frac{\gamma\left(\eta_1,\eta_2 t_1\right)}{\Gamma\left(\eta_1\right)}=p_{0,t_1},\\
	\frac{\gamma\left(\eta_1,\eta_2 t_2\right)-\gamma\left(\eta_1,\eta_2  t_1\right)}{\Gamma\left(\eta_1\right)}=p_{t_1,t_2}.
\end{dcases}
\end{equation*}

For values of $t_1,t_2$ that mirror those from table \ref{table:derivatives}, and for some arbitrary choices of the underlying probabilities, we obtain the results in table \ref{table:size-biased}. These results are purely for exemplification, and the approach that we outline in this section can be used to calculate $\boldsymbol{\eta}$ for any partition of $\left(0,\infty\right)$, any associated probabilities, and any distributional choice of $p\left(\alpha\mid\boldsymbol{\eta}\right)$. 

Unconditionally, we have that:
\begin{equation*}
p\left(w_1,w_2\right)=\int_0^\infty\alpha^2\ \frac{\left(1-w_1-w_2\right)^{\alpha-1}}{1-w_1}\ \textrm{d}p\left(\alpha\mid\boldsymbol{\eta}\right),
\end{equation*}
which, when $\alpha\sim\textrm{Ga}\left(a,b\right)$, leads to the following analytic expressions (see \ref{section:appendix5}):
\begin{align*}
p\left(w_1,w_2\right)&=a\left(a+1\right)b^a \ \frac{\left(b-\log\left(1-w_1-w_2\right)\right)^{-a-2}}{\left(1-w_1\right)\left(1-w_1-w_2\right)},\\
p\left(w_1\right)&=ab^a\ \frac{\left(b-\log\left(1-w_1\right)\right)^{-a-1}}{\left(1-w_1\right)},
\end{align*}
which can potentially be used for further analytical considerations, when setting $\left(a,b\right)$.

\begin{figure}[htb]
\centering
\includegraphics[width=0.98\textwidth]{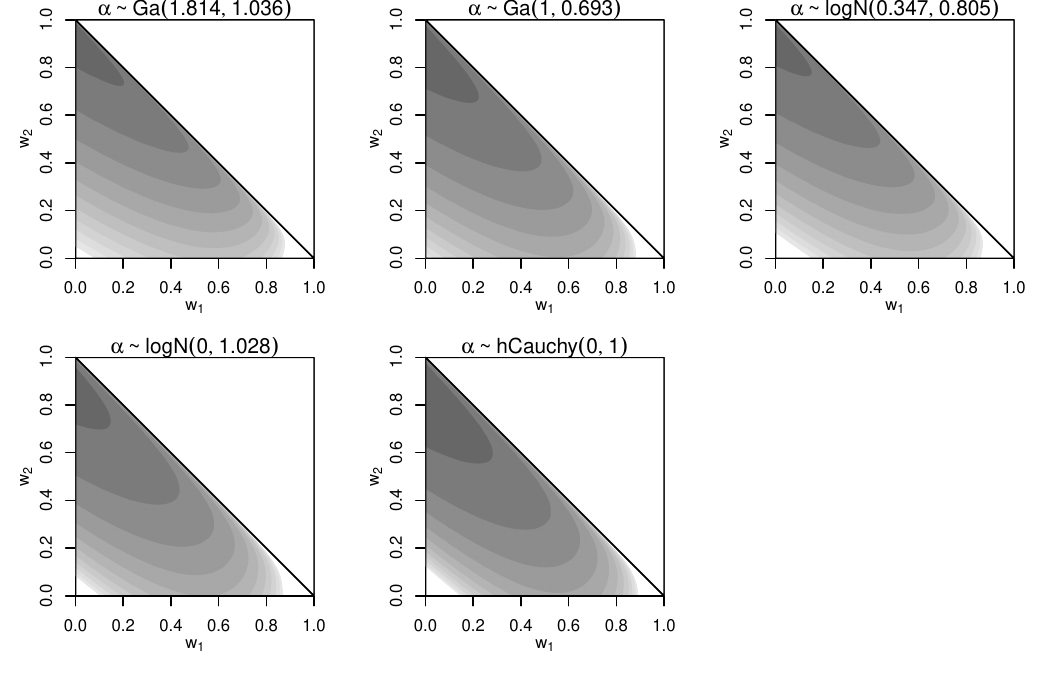}
\caption{Sample-size-independent approach, size-biased. Joint probability distribution $p\left(w_1,w_2\right)$, for the distributional choices of $\alpha$ identified in table \ref{table:size-biased}. Darker colours indicate smaller values.}
\label{figure:SB_mixed}
\vspace{1cm}
\end{figure}

The plots in Figure \ref{figure:SB_mixed} confirm that the joint unconditional distribution reflects a mix of the characteristics of the conditional joint distributions, in that the probability is amassed at the vertices $\left(0,1\right)$ and $\left(1,1\right)$, and along the edge that connects $\left(1,0\right)$ and $\left(0,0\right)$. 

\begin{table}[tb]
\centering
\begin{tabular}{lccccc}
	Distribution & $p\left(0<\alpha<1\right)$ & $p\left(1<\alpha<2\right)$ & $p\left(\alpha>2\right)$ & $\eta_1$ & $\eta_2$ \\ \hline 
	Gamma & $1/3$ & $1/3$ & $1/3$ & 1.814 & 1.036\\
	Gamma & $1/2$ & $1/4$ & $1/4$ & 1.000 & 0.693\\
	Lognormal & $1/3$ & $1/3$ & $1/3$ & 0.347 & 0.805\\
	Lognormal & $1/2$ & $1/4$ & $1/4$ & 0.000 & 1.028\\
	Half-Cauchy & 1/2 & \multicolumn{2}{c}{$p\left(\alpha>1\right)=1/2$} & 0.000 & 1.000
\end{tabular}
\caption{Sample-size-independent approach, size-biased. A selection of results for different distributional choices of $\alpha$, and for different choices of $p\left(0<\alpha<1\right)$, $p\left(1<\alpha<2\right)$, $p\left(\alpha>2\right)$.}
\label{table:size-biased}	
\end{table}

\subsection{Ranked weights}\label{section:ranked}
\begin{figure}[htb]
\centering
\includegraphics[width=0.98\textwidth]{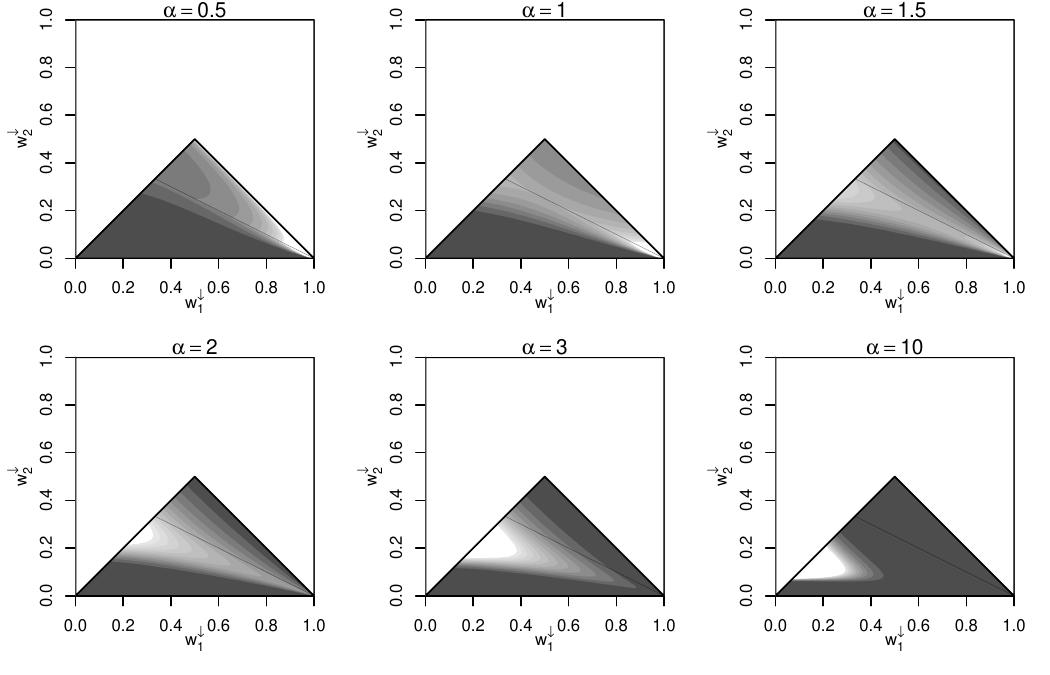}
\caption{Sample-size-independent approach, ranked. Joint probability distribution $p\left(w^{\shortdownarrow}_1,w^{\shortdownarrow}_2\mid\alpha\right)$, for different values of $\alpha$. The top right triangle identifies $A \subset E$ (see equation \ref{eq:ranked2}). Darker colours indicate smaller values.}
\label{figure:ranked_joint}
\end{figure}

The same approach from section \ref{sec:size} can also be applied to ranked stick weights, in the Poisson-Dirichlet distribution representation (see section \ref{section:pdd}). We re-write equation \ref{eq:watterson} as follows:
\begin{equation}\label{eq:ranked1}
p\left(w_1^\shortdownarrow,\ldots, w_r^\shortdownarrow\mid\alpha\right)=\alpha^r \frac{\left(1-w_1^\shortdownarrow-\ldots-w_r^\shortdownarrow\right)^{\alpha-1}}{w_1^\shortdownarrow\cdots w_r^\shortdownarrow}\ F_{\alpha}\left(\frac{w_r^\shortdownarrow}{1-w_1^\shortdownarrow-\ldots-w_r^\shortdownarrow}\right),
\end{equation}
with $F_\alpha\left(x\right):=p\left(w_1^\shortdownarrow\leq x\mid\alpha\right)$, which can be obtained by simulation. In the bivariate case, equation \ref{eq:ranked1} becomes:
\begin{equation}\label{eq:ranked2}
p\left(w_1^\shortdownarrow,w_2^\shortdownarrow\mid\alpha\right)=\alpha^2 \frac{\left(1-w_1^\shortdownarrow-w_2^\shortdownarrow\right)^{\alpha-1}}{w_1^\shortdownarrow w_2^\shortdownarrow}\ F_{\alpha}\left(\frac{w_2^\shortdownarrow}{1-w_1^\shortdownarrow-w_2^\shortdownarrow}\right),
\end{equation}
which is defined on 
\begin{equation*}
E=\left\{\left(w_1^\shortdownarrow,w_2^\shortdownarrow\right):\left(w_1^\shortdownarrow+w_2^\shortdownarrow<1\right)\ \land \ \left(w_2^\shortdownarrow<w_1^\shortdownarrow \right)\right\}.
\end{equation*}
Since $F_\alpha$ is a cumulative probability distribution function, we have that
\begin{equation*}
F_{\alpha}\left(\frac{w_2^\shortdownarrow}{1-w_1^\shortdownarrow-w_2^\shortdownarrow}\right)=1,\quad \textrm{for } w_2^\shortdownarrow\geq \frac{1}{2}-\frac{w_1^\shortdownarrow}{2}.
\end{equation*}
As such, the restriction of equation \ref{eq:ranked2} to 
\begin{equation*}
A=\left\{\left(w_1^\shortdownarrow,w_2^\shortdownarrow\right):\left(w_1^\shortdownarrow+w_2^\shortdownarrow<1\right)\ \land \ \left(w_2^\shortdownarrow<w_1^\shortdownarrow\right) \ \land \ \left(w_2^\shortdownarrow\geq \frac{1}{2}-\frac{1}{2} w_1^\shortdownarrow\right) \right\}
\end{equation*}
yields
\begin{equation}\label{eq:ranked3}
p\mid_A\left(w_1^\shortdownarrow,w_2^\shortdownarrow\mid\alpha\right)=\alpha^2 \frac{\left(1-w_1^\shortdownarrow-w_2^\shortdownarrow\right)^{\alpha-1}}{w_1^\shortdownarrow w_2^\shortdownarrow},
\end{equation}
whose partial derivatives are easier to study than those of equation \ref{eq:ranked2}, and which bring some insights. In particular:
\begin{align*}
\frac{\partial p\mid_A \left(w_1^\shortdownarrow,w_2^\shortdownarrow\mid\alpha\right)}{\partial w_1^\shortdownarrow}&=\alpha^2 \frac{\left(1-w_1^\shortdownarrow-w_2^\shortdownarrow\right)^{\alpha-2}}{w_1^{\shortdownarrow 2}w_2^\shortdownarrow}\left(\left(2-\alpha\right)w_1^\shortdownarrow+w_2^\shortdownarrow-1\right),\\
\frac{\partial p\mid_A \left(w_1^\shortdownarrow,w_2^\shortdownarrow\mid\alpha\right)}{\partial w_2^\shortdownarrow}&=\alpha^2 \frac{\left(1-w_1^\shortdownarrow-w_2^\shortdownarrow\right)^{\alpha-2}}{w_1^\shortdownarrow w_2^{\shortdownarrow 2}}\left(\left(2-\alpha\right)w_2^\shortdownarrow+w_1^\shortdownarrow-1\right),
\end{align*}
which leads to the considerations in table \ref{table:derivativesRanked}. Following the same approach we used in section \ref{sec:size}, we also plot in Figure \ref{figure:ranked_joint}  the joint bivariate distribution of $p\big(w_1^\shortdownarrow,w_2^\shortdownarrow\mid\alpha\big)$ for key values of $\alpha$, to spot any further visible patterns in the behaviour of $p\left(w_1^\shortdownarrow,w_2^\shortdownarrow\mid\alpha\right)$ as $\alpha$ varies. 

\begin{table}[tb]
\centering
\begin{tabular}{lll}
	$\alpha$ & $p\mid_A\left(w_1^\shortdownarrow\mid w_2^\shortdownarrow,\alpha\right)$ & $p\mid_A\left(w_2^\shortdownarrow\mid w_1^\shortdownarrow,\alpha\right)$ \\ \hline
	$0<\alpha<1$ & convex; min. at $w_1^\shortdownarrow=\frac{1-w_2^\shortdownarrow}{2-\alpha}$ 
	& convex; min. at $w_2^\shortdownarrow=\frac{1-w_1^\shortdownarrow}{2-\alpha}$\\ 
	$\alpha\geq1$ & decreasing & decreasing
\end{tabular}
\caption{Sample-size-independent approach, ranked. Behaviour of $w_1^\shortdownarrow\mid w_2^\shortdownarrow,\alpha$ and $w_2^\shortdownarrow\mid w_1^\shortdownarrow,\alpha$, for different values of $\alpha$, over $A$.}
\label{table:derivativesRanked}	
\end{table}

\begin{figure}[htb]
\centering
\includegraphics[width=0.98\textwidth]{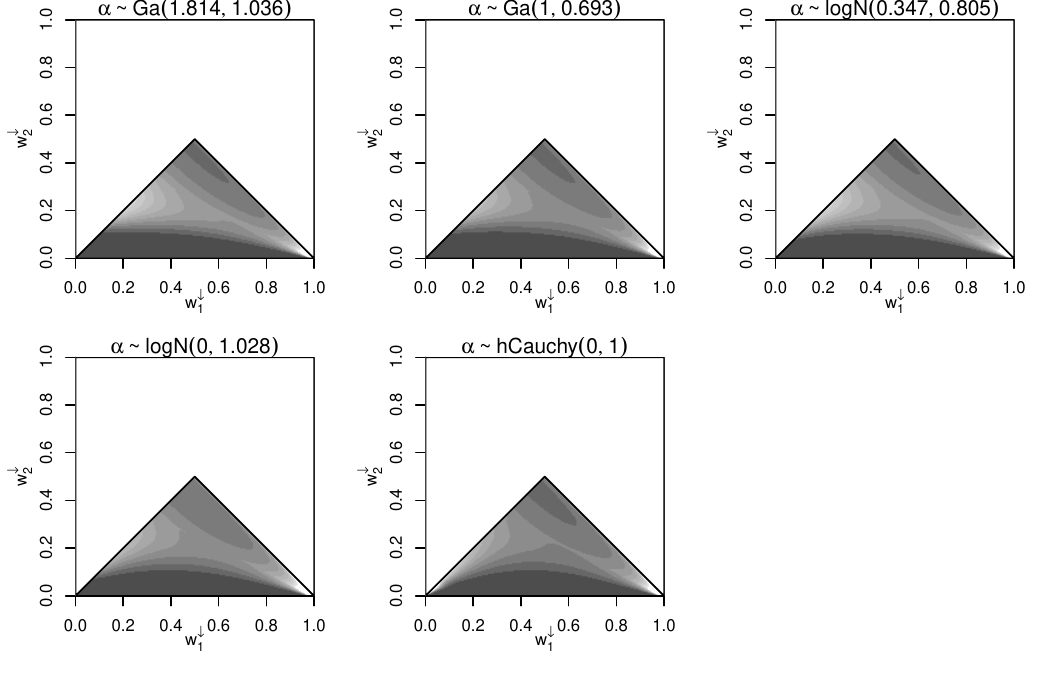}
\caption{Sample-size-independent approach, ranked. Joint probability distribution $p\left(w^{\shortdownarrow}_1,w^{\shortdownarrow}_2\right)$, for the distributional choices of $\alpha$ identified in Table \ref{table:size-biased}. Darker colours indicate smaller values.}
\label{figure:ranked_joint_random}
\end{figure}

The process for choosing the probability distribution $p\left(\alpha\mid\boldsymbol{\eta}\right)$ and its parameters is analogous to the one in section \ref{sec:size}. In Figure \ref{figure:ranked_joint_random}, we plot $p\left(w_1^\shortdownarrow,w_2^\shortdownarrow\right)$ for different distributional choices of $p\left(\alpha\mid\boldsymbol{\eta}\right)$, to show the impact of marginalising $\alpha$ out.

\section{Discussion}\label{section:comparison}
In section \ref{section:3} we studied the behaviour of SSD, quasi-degenerate and improper priors in view of the distribution that they induce on $K_n$, while in \ref{section:SSI} we assessed SSI priors with respect to their implied stick-breaking distribution. Here we do the opposite: we cross-check how SSD and SSI priors behave in relation to each other's driving metric. For the sake of brevity, we only discuss size-biased stick breaking weights; conclusions with respect to ranked weights are similar.

\begin{figure}[htb]
\centering
\includegraphics[width=0.98\textwidth]{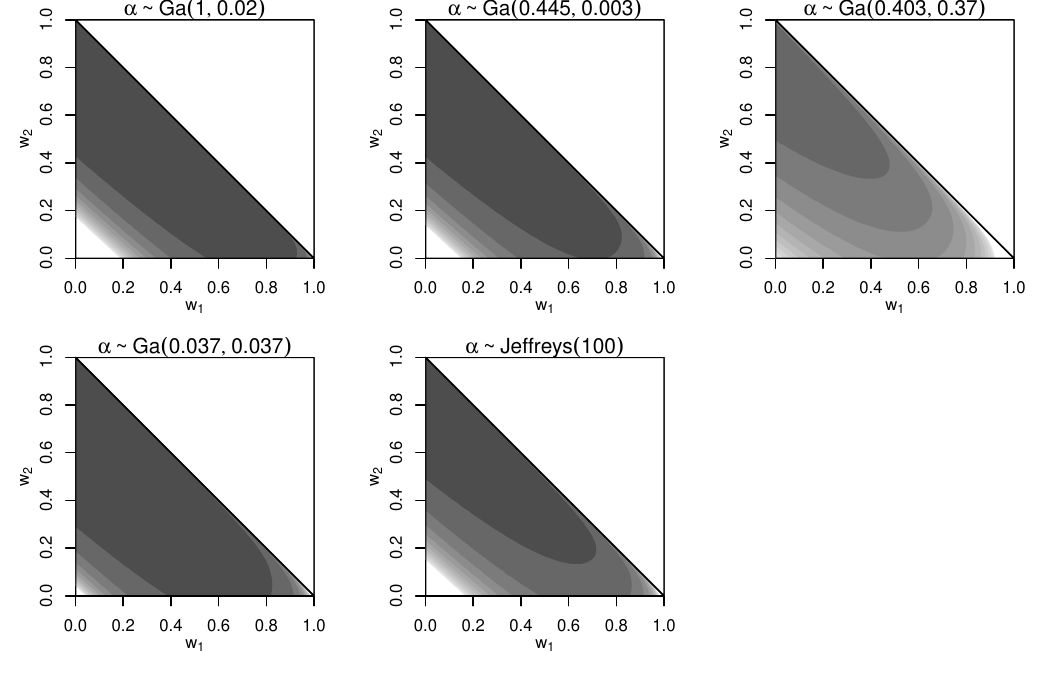}
\caption{Size-biased joint probability distribution $p\left(w_1,w_2\right)$ underlying the $K_n$-diffuse, DORO, SCAL, quasi-degenerate and Jeffreys' priors for $n=100$. Darker colours indicate smaller values.}
\label{figure:SB_mixed2}
\vspace{20pt}
\includegraphics[width=0.93\textwidth]{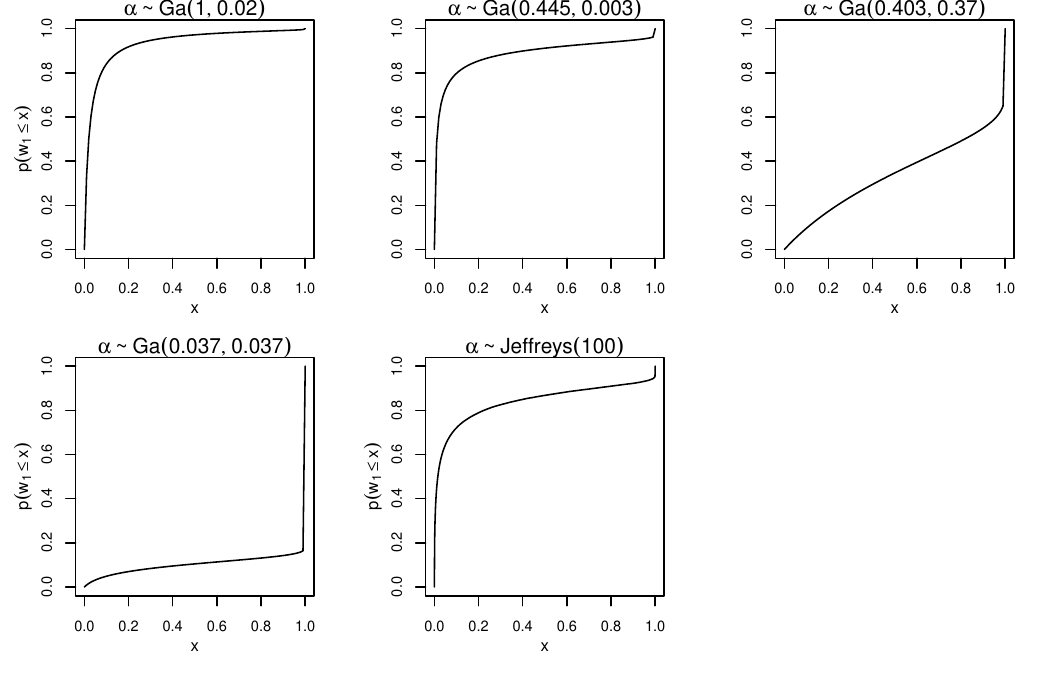}
\caption{Size-biased cumulative probability distributions of $w_1$ underlying the $K_n$-diffuse, DORO, SCAL, quasi-degenerate and Jeffreys' priors for $n=100$.}
\label{figure:SSD_ranked}
\end{figure}

\begin{figure}[htb]
\centering
\includegraphics[width=0.98\textwidth]{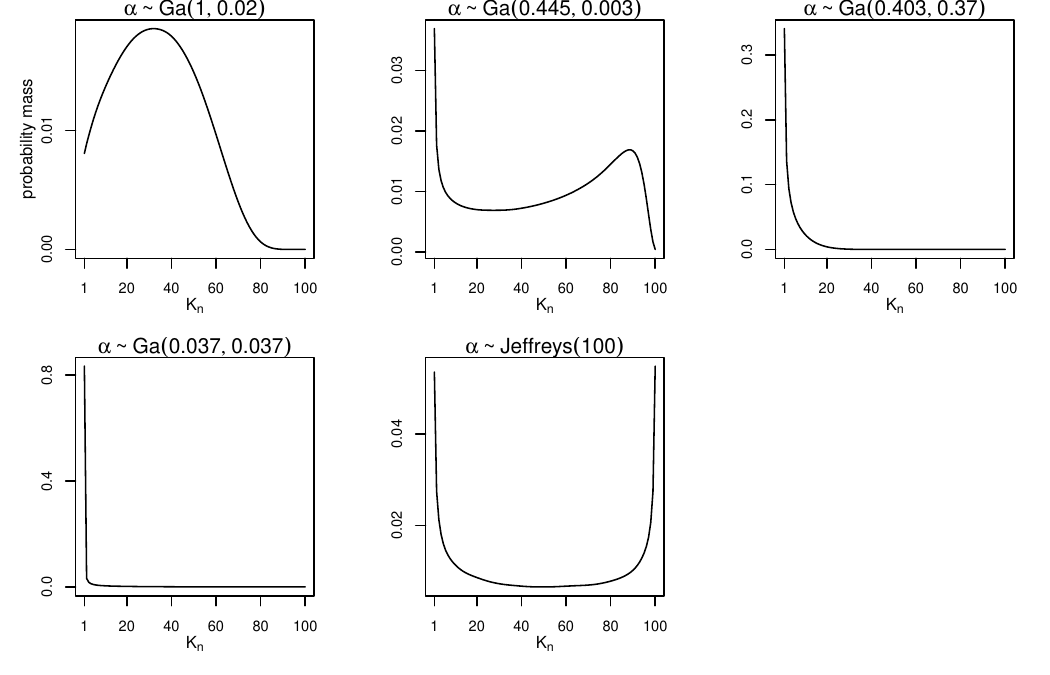}
\caption{Prior distributions $p\left(K_n\right)$ induced by $p\left(\alpha\right)$, for the $K_n$-diffuse, DORO, SCAL, quasi-degenerate and Jeffreys' priors for $n=100$.}
\label{figure:prior03}
\end{figure}

We observe that the behaviour of the $K_n$-diffuse, the DORO, the quasi-degenerate\footnote{We parameterized the quasi-degenerate prior as a $\textrm{Ga}\left(0.403,0.370\right)$, to mirror the result from the approximation method of SCAL, although we could have used any smaller value of $\left(a,b\right)$. Conclusions would not materially change.} and Jeffreys' priors with respect to $p\left(w_1,w_2\right)$ (Figure \ref{figure:SB_mixed2}) is markedly different from the behaviour of SSI (Figure \ref{figure:SB_mixed}), and more extreme, because probability in SSD is mostly concentrated at $\left(0,0\right)$ or $\left(1,0\right)$, while in SSI it is more spread out. SCAL is closer to SSI in this respect. By looking at \ref{figure:SSD_ranked} too, we conclude that the $K_n$-diffuse, the DORO and Jeffreys' priors would likely attract posterior estimates of $\left(w_1,w_2\right)$ towards $\left(0,0\right)$, while the quasi-degenerate prior would attract them towards $\left(1,0\right)$. We also note that neither SCAL nor the quasi-degenerate priors are diffuse in $K_n$ (Figure \ref{figure:prior03}).

\begin{figure}[htb]
\centering
\includegraphics[width=0.98\textwidth]{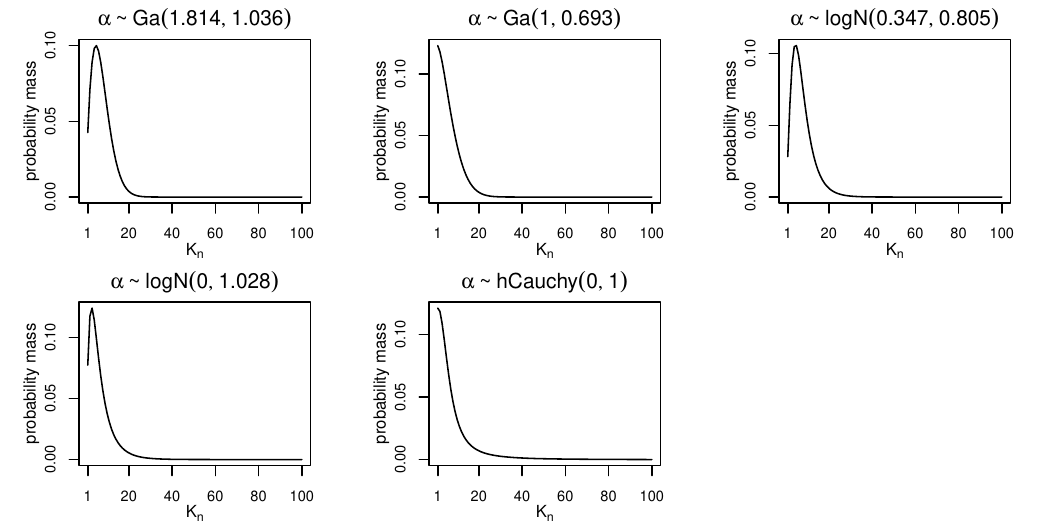}
\caption{Prior distributions $p\left(K_n\right)$ induced by $p\left(\alpha\right)$, for $n=100$, for the distributional choices of $\alpha$ identified in Table \ref{table:size-biased}.}
\label{figure:prior02}
\end{figure}

Conversely, with respect to SSI, we consider the five test cases identified in Table \ref{table:size-biased}, and we plot their implied distribution of $K_n$ in Figure \ref{figure:prior02}. We observe that their implied $p\left(K_n\right)$ is concentrated over values that are small, relative to $n$. We highlight that these five cases are just examples, as one's prior information to be reflected with SSI may well be entirely different from the cases in table \ref{table:size-biased}.

We more generally conclude that choices of $p\left(\alpha\mid\boldsymbol{\eta}\right)$ that are diffuse in $K_n$ are not necessarily diffuse in $\left(w_1,w_2\right)$, and vice-versa. As such, users should assess whether the SSD or SSI prior approach is more suitable with respect to their particular  problem at hand; using both $p\left(K_n\right)$ and $p\left(w_1,w_2\right)$ as a reference to set $p\left(\alpha\mid\boldsymbol{\eta}\right)$ could also be an option, if $p\left(K_n\right)$ is relevant to the application. We argue that SSI priors still hold a number of advantages over SSD, as we summarize in Section \ref{section:conclusions}. 

\section{Case study: multiple DPMs}\label{section:5}
In this section we point to a setting where, by construction, the sample-size-dependent approach to choosing $p\left(\alpha\mid\boldsymbol{\eta}\right)$ is inapplicable, while the sample-size-independent approach can still be used. This was inspired by \cite[figure 5.1]{muller_nonparametric_2013}.

Consider a partially exchangeable setting of $J$ groups with $n_j$ observations in group $j$, where $j=1,\ldots,J$ and the groups share the same common underlying precision parameter $\alpha$. This framework, while extremely simple, allows information to be borrowed between groups through their dependence on a common $\alpha$. 

We denote the observations by $y_{i,j}, \ i=1,\ldots,n_j, \ j=1,\ldots,J$, and the random number of clusters in group $j$ by $K_{n,j}$. The model is:
\begin{align*}
&y_{i,j}\mid\theta_{i,j}\sim p\left(y_{i,j}\mid\theta_{i,j}\right)\\
&\theta_{i,j}\mid G_j\sim G_j\\
&G_j\mid\alpha\sim \textrm{DP}\left(\alpha,G_{0j}\right)\\
&\alpha\sim p\left(\alpha\mid\boldsymbol{\eta}\right).
\end{align*}

It is clear that, by construction, multiple $K_{n,j}$ are involved, hence there is no unique probability distribution for $K_n$ and no unique sample size $n$ that can be used as a target for the sample-size-dependent approaches that we have described. Conversely, our sample-size-independent approach is still applicable.

A different example of a set-up that is similarly unsuitable for SSD priors due to the lack of a unique sample size is that of online inferential algorithms for streaming data, because the sample size changes continuously as more data is collected.

\section{Conclusions}\label{section:conclusions}
In section \ref{section:3} we highlighted the limitations of previous approaches. Their limitations include:
\begin{itemize}
\item Dependence on $n$. All SSD priors are only optimal for one particular value of $n$, and they need to be updated as $n$ varies;
\item Unmet assumptions. DORO minimises the Kullback-Leibler distance between $p\left(K_n\right)$ and the discrete uniform distribution, but the discrete uniform target can never be attained, as $p\left(K_n\right)$ converges to a negative binomial for $n\rightarrow \infty$ (see section \ref{section:dorazio}). The assumption that a non-informative prior would be reflected in a discrete uniform distribution induced on $K_n$ is not supported by Jeffreys' prior either, which has a very different shape from discrete uniform (see Figure \ref{figure:jeffreys-k});
\item Asymptotic breakdown. The approximation suggested by SCAL in relation to their test case, as an alternative to their fully-fledged approach, is only valid over a narrow range of values of $n$; for moderately large $n$, it results in $\alpha\sim\textrm{Ga}\left(a,b\right)$, $\left(a,b\right)\approx \left(0,0\right)$. This is undesirable, as it implies $p\left(K_n=1\right)\approx 1$, which negates the reason for using a nonparametric model in the first place (see sections \ref{section:murugiah}, \ref{section:degenerate});
\item Over-informativeness. Although $K_n$-diffuse and DORO priors are diffuse in $K_n$, they are very informative with respect to $\left(w_1,w_2\right)$ and $\left(w_1^\shortdownarrow,w_2^\shortdownarrow\right)$ (see section \ref{section:comparison});
\item Inapplicability. SSD priors are inapplicable when multiple DPMs with different sample sizes share the same $\alpha$ (see section \ref{section:5}); SSD priors are not well suited either to cases where the sample size is allowed to grow. 
\item Incompatibility. Quasi-degenerate priors are not compatible with the notion of multiple clusters, hence they defeat the purpose of using a DPM in the first place (see \ref{section:degenerate});
\item Impropriety. Improper priors can lead to an improper posterior $p\left(\alpha\mid\boldsymbol{y}\right)$ as well as to an improper prior induced on $K_n$ (see \ref{section:improper2}).  
\end{itemize}

In \ref{section:SSI}, we introduced a new approach, based on the appraisal of the implied joint distribution of the stick-breaking weights (either in size-biased order, or ranked). This is equivalent to thinking in terms of the asymptotic relative cluster sizes for $n\rightarrow\infty$. Our approach does not suffer from any of the aforementioned limitations, although we acknowledge that SSD priors still have a place, depending on the specific nature of the problem at hand, and that potentially both $p\left(K_n\right)$ and the stick-breaking weights could be concurrently leveraged to inform one's choice of $p\left(\alpha\mid\boldsymbol{\eta}\right)$.

Out of the two alternatives that we propose (size-biased weights, and ranked weights), the one that is based on $p\left(w_1,w_2\right)$ appears to be the easiest to interpret, compute, and evaluate, due to the distinctive behaviour of $p\left(w_1,w_2\mid\alpha\right)$ for various levels of $\alpha$ (Figure \ref{figure:SB_conditional}) and to the availability of simpler analytical formulae. Quantitative measurements can be carried out, as exemplified in Table \ref{table:size-biased}, to determine how to mix precisely over those behaviours. A simple exact formula is also available when $\alpha\sim\textrm{Exp}\left(\eta\right)$ (see equation \ref{eq:explicit}). We envisage use of SSI priors by practitioners as a valid alternative to SSD priors, not only because of the principle that underlies them (which has general appeal in all applications) but also because they allow for two applications that would not otherwise be feasible with SSD priors. One is the area of online learning algorithms, where the focus is to obtain an end-to-end inferential algorithm for streaming data that allows updates to the posterior estimates as the size of the data grows: in such applications, a prior that does not depend on the sample size is clearly very desirable. DPM models are also particularly well suited to streaming data as they allow the number of clusters to grow as the size of the data set increases. The second is the case of multiple DPM models that share a common underlying precision parameter, which allows the borrowing of strength between models and which is not feasible with traditional sample-size-dependent priors because there is no single sample size with which to parameterize the prior (see section \ref{section:5}).

\appendix
\section{Unconditional size-biased weight distribution}\label{section:appendix5}
When $\alpha\sim\textrm{Ga}\left(a,b\right)$, the joint distribution of the first $H$ size-biased weights of a stick-breaking process has a simple closed analytical formula, once $\alpha$ is integrated out of equation \ref{eq:connorMosimann}:
\begin{align}\nonumber
	p\left(w_1,\ldots,w_H\right)&=\int_0^\infty \frac{\alpha^H\left(1-w_1-\ldots-w_H\right)^{\alpha-1}}{\left(1-w_1\right)\ldots\left(1-w_1-\ldots-w_{H-1}\right)}dp\left(\alpha\right)\\
	&=\frac{\Gamma\left(a+H\right)}{\Gamma\left(a\right)}\ b^a\ \frac{\left[b-\log\left(1-w_1-\ldots-w_H\right)\right]^{-a-H}}{\left(1-w_1\right)\ldots\left(1-w_1-\ldots-w_H\right)}\\
	&\propto \frac{\left[b-\log\left(1-w_1-\ldots-w_H\right)\right]^{-a-H}}{\left(1-w_1\right)\left(1-w_1-\ldots-w_H\right)}.\label{eq:newDistr}
\end{align}
The cumulative probability distribution of $w_1$ also has a simple explicit analytical representation:
\begin{equation*}
	p\left(w_1\leq x\right)=\int_0^x p\left(w_1\right)\textrm{d}w_1=1-\left(\frac{b}{b-\log\left(1-x\right)}\right)^a.
\end{equation*}

\section{Sampling from Jeffreys' prior}\label{section:jeffreysMC}
As Jeffreys' prior for the Dirichlet process is a non-standard distribution, there is value in discussing how to sample from it. Like any other distribution for which there are no widely known sampling methods, one can resort to any of the general-purpose approaches described, for example, in \cite{christian_monte_2004}, such as the accept-reject method, importance sampling, the 2-d slice sampler or Metropolis-Hastings algorithms. 

The accept-reject algorithm requires that a constant $M$ exists such that the target distribution $f$ and the proposal distribution $g$ meet the condition $\frac{f\left(\alpha\right)}{g\left(\alpha\right)}\leq M$, for all values of $\alpha$. This can be achieved by using Jeffreys' prior for $n=2$ (which has an analytical solution, as per equations \ref{eq:Jdensity} and \ref{eq:Jcumulate}) as the proposal distribution $g$; easy calculations lead then to
\begin{equation*}
\frac{f\left(\alpha\right)}{g\left(\alpha\right)}=\sqrt{1+2\frac{\left(\alpha+1\right)^2}{\left(\alpha+2\right)^2}+\ldots+\left(n-1\right)\frac{\left(\alpha+1\right)^2}{\left(\alpha+n-1\right)^2}}\leq M =\sqrt{\sum_{i=1}^{n-1} i}.
\end{equation*}
Once $M$ is determined, the steps to the algorithm are:
\begin{itemize}
\item generate $X$ from the proposal distribution $g$, and generate $U\sim\textrm{Unif}\left(0,1\right)$;
\item accept the value above if $U\leq f\left(X\right)/\left(M g\left(X\right)\right)$; return to the step above if otherwise.
\end{itemize}
We also implemented, to compare:
\begin{itemize}
\item the 2d slice sampler
\item the independence Metropolis-Hastings algorithm with a Jeffreys' prior distribution ($n=2)$;
\item the random-walk Metropolis algorithm with half-Cauchy and normal proposal distributions;
\end{itemize}
and found the convergence speed to be fastest for the 2d slice sampler, followed by the independence Metropolis-Hastings algorithm with Jeffreys' prior distribution and $n=2$, followed again by the random-walk Metropolis with half-Cauchy proposal. By comparison, \cite{rodriguez2013jeffreys} use random-walk Metropolis-Hastings with a normal proposal distribution.

\section{Propriety of the posterior induced by Jeffreys' prior on \texorpdfstring{$\alpha$}{alpha}}\label{appendix:proprietyA}
Jeffreys' prior induces a proper posterior $p\left(\alpha\mid K_n=k\right)$. Given 
\begin{align*}
p\left(\alpha\mid K_n=k\right)&\propto p\left(K_n=k\mid\alpha\right) p\left(\alpha\right)\\
&\propto \alpha^k\frac{\Gamma\left(\alpha\right)}{\Gamma\left(\alpha+n\right)}\sqrt{\frac{1}{\alpha}\sum_{i=1}^n\frac{i-1}{\left(\alpha+i-1\right)^2}},
\end{align*}
its limit for $\alpha\rightarrow 0$ is
\begin{equation*}
\lim_{\alpha\rightarrow0} p\left(\alpha\mid K_n=k\right)=\alpha^{k-1}\frac{\Gamma\left(1\right)}{\Gamma\left(n\right)}\alpha^{-\frac{1}{2}}\sqrt{\sum_{i=1}^n\frac{1}{i-1}}\approx \alpha^{k-\frac{3}{2}},
\end{equation*}
which is convergent over $\left(0,1\right)$ (limit comparison test). Its limit for $\alpha\rightarrow\infty$ is
\begin{equation*}
\lim_{\alpha\rightarrow\infty} p\left(\alpha\mid K_n=k\right)=\alpha^{k-n-\frac{3}{2}},
\end{equation*}
which is proper over $\left(t,\infty\right)$, for $t\geq1$.

\section{Propriety of the prior induced by Jeffreys' prior on \texorpdfstring{$K_n$}{Kn}}\label{appendix:proprietyK}
Recall from equation \ref{eq:K} the expression for $p\left(K_n=k\mid\alpha\right)$. The prior it induces on $K_n$ is
\begin{align*}
p\left(K_n=k\right)&=s_{n,k}\int_0^\infty \alpha^k\frac{\Gamma\left(\alpha\right)}{\Gamma\left(\alpha+n\right)}\sqrt{\frac{1}{\alpha}\sum_{i=1}^n\frac{i-1}{\left(\alpha+i-1\right)^2}}\:\textrm{d}\alpha\\
&\propto s_{n,k}\int_0^\infty \frac{\Gamma\left(\alpha\right)}{\Gamma\left(\alpha+n\right)}\alpha^{k-\frac{1}{2}}\sqrt{\sum_{i=1}^n\frac{i-1}{\left(\alpha+i-1\right)^2}}\:\textrm{d}\alpha.
\end{align*}
Denoting the integrand by $g\left(\alpha\right)$,  
\begin{align*}
\lim_{\alpha\rightarrow 0} g\left(\alpha\right)&=\lim_{\alpha\rightarrow 0}\frac{\Gamma\left(\alpha+1\right)}{\Gamma\left(\alpha+n\right)}\alpha^{k-\frac{3}{2}}\sqrt{\sum_{i=1}^n \frac{i-1}{\left(\alpha+i-1\right)^2}}\\
&=\lim_{\alpha\rightarrow 0}\frac{\alpha^{k-\frac{3}{2}}}{\Gamma\left(n\right)}\sqrt{\sum_{i=1}^n\frac{1}{i-1}}\\
&\propto \alpha^{k-\frac{3}{2}},
\end{align*}
which converges over $\left(0,1\right)$, and
\begin{equation*}
\lim_{\alpha\rightarrow \infty} g\left(\alpha\right)\propto \alpha^{k-\frac{1}{2}-n-1},
\end{equation*}
which converges over $\left(1,\infty\right)$.
\addcontentsline{toc}{chapter}{Bibliography}
\nocite{}
\bibliographystyle{elsarticle-harv}
\bibliography{bibliography}

\end{document}